# Low-cost Efficient Wireless Intelligent Sensor (LEWIS) for Engineering, Research, and Education

*Mahsa Sanei [a], Solomon Atcitty [b], Fernando Moreu [a*]*

a. *Department Civil, Construction, and Environmental Engineering, University of New Mexico, Albuquerque, NM 87131-0001, USA*

b. *Department of Mechanical Engineering, University of New Mexico, Albuquerque, NM 87131-0001, USA*

**Corresponding author's email address**
*Email: F. Moreu, fmoreu@unm.edu*

**Abstract**

Sensors have the capability of collecting engineering data and quantifying environmental changes, activities, or phenomena. Civil engineers are not prone to designing, fabricating, and installing sensors for their day-to-day decisions in specialized professionals given their lack of knowledge in sensor technology. Therefore, the vision of smart cities equipped with sensors informing decisions has not been realized to date. Additionally, the cost associated with data acquisition systems, laboratories, and experiments restricts access to sensors as learning tools for wider audiences. Recently, sensors are becoming a new tool in education and training, giving learners real-time information that can reinforce their confidence and understanding of scientific or engineering new concepts. However, the electrical components and required computer knowledge associated with sensors are still a challenge for civil engineers without such a background. If sensing technology costs, complexity, and use are simplified, sensors could be tamed by civil engineering students. The researcher developed, fabricated, and tested an efficient low-cost wireless intelligent sensor (LEWIS) aimed at education and research, named LEWIS1. This platform is directed at learners connected with a cable to the computer but has the same concepts and capabilities as the wireless version. The content of this paper describes the hardware and software architecture of the first prototype and their use, as well as the proposed new LEWIS1 (LEWIS1-β) that simplifies both hardware and software, in addition to the user interfaces. The capability of the proposed sensor is compared with an accurate commercial PCB sensor through a set of experiments. The later part of this paper demonstrates applications and examples of outreach efforts and suggests the adoption of LEWIS1-β as a new tool for education and research. The authors also investigated the number of activities and sensor building workshops that has been done since 2015 using the LEWIS sensor which shows an ascending trend of different profession's people excitement to involve and learn the sensor fabrication.

## 1. Introduction

The Vision for Civil Engineering in 2025 is a strategic plan presented by the American Society of Civil Engineers (ASCE). This strategic plan demonstrates the worldwide vision for the civil engineering future and delineates the importance of engaging more people in the field of civil engineering [1]. According to ASCE 2025 vision, engineers rely on using real-time access to databases, sensors, analytical tools, and other sophisticated technology to help them make informed decisions. Smart cities (sensors and real-time on-board diagnostics system) has contributed to this rapid advance change and the integration of high-quality technology into the life of structures. [2]. Prediction times have significantly risen due to real-time monitoring, acquisition, collecting data, storage, and modeling, which has also made informed-decisions possible. Hence, in the future vision of ASCE, there should be a plan that not only emphasizes using the sensor broadly but also consider transferring the knowledge and application of this technology to the new generation in a simple way. An engineer/student can apply knowledge or abilities they have mastered in one environment to another situation through a transfer cognitive process. Comprehension of the learner helps them understand how their knowledge is relevant and can be successfully applied outside the original learning context [3]. Therefore, the knowledge of sensors should be brought up in a broader context to be transferable and applicable in a variety of applications.

Structural Health Monitoring (SHM) provides with algorithms and resources to stakeholders with recurring evaluations of the integrity and safety of civil infrastructure [4]. Civil engineers can benefit from SHM by monitoring structural data and making informed-decisions, which are essential for the safe functioning of infrastructure [5]. Sensors are extensively utilized in SHM systems to quantify structural responses over time (displacement, strain, rotation, temperature, humidity, acceleration, etc.) and consequently identify any unusual issues that can have detrimental effects, such structural damage [6]. The information recorded from sensors is used to estimate structural performance [7]. Assessment of the structures and their health status is necessary to diminish repair costs and maintenance and ultimately prove infrastructure safety [8]. Concepts of "smart cities" have been put up as a way to use technology-based solutions to alleviate infrastructure issues and enhance infrastructure performance. The application of smart technology such as sensors for infrastructure systems has been identified a solution which highlights new responsibilities for civil engineers that can help them develop applications for smart cities [9].

The demand for effective and affordable monitoring systems such as sensors keeps rising. Accelerometer sensors are found en-masse in a variety of prices and specifications through online markets and vendors. Many of these are effective in both cost and accuracy. The low-cost sensors provide a number of advantages over traditional instrumentation, including a lower unit price, a smaller size, portability, and the capacity to capture acceleration in three different axes [10]. However, a thorough examination of these information technologies, their uses over the whole life cycle of infrastructure, and systematic analysis and debate of them are lacking.

To date, different researchers have used or developed low-cost sensors in many applications. Due to the rapid development of commercial wireless and mobile computer technologies, the hardware capabilities of sensing unit designs continue to advance. Parallel to the advancements in functionality, wireless sensors' price and form factor have decreased. The advantage of wireless monitoring systems with high sensor density is a direct result of the low price of wireless sensing modules. Dense-sensor deployments give more information about how the structure behaves at component-level length scales, producing empirical response data that can be used to more precisely identify damage [11]. However, these sensors are intimidating for layman users as they require deep knowledge of circuitry and electrical components.

The variety of applications for information sensing technology has significantly increased as the price of sensor equipment has decreased in recent years [12]. Research in academia is expanding on the use of sensor technology in different areas. Weng et al. developed and validated a real-time health monitoring system using integrated wireless network system for civil constructions. The suggested method allows for the simultaneous collection and analysis of data from a number of wireless sensing units that can operate as a set of analog sensors in real-time. The sensor signals are enhanced by incorporating low-cost signal conditioning circuits. Extensive lab and field tests support the viability and dependability of this integrated wireless SHM network system [13]. Girolami et al. instrumented a structure by dispersing several synchronized low-cost LIS344ALH accelerometers alongside with it to develop a system which is capable of obtaining the structural modal. Although the proposed system of synchronized sampling from different nodes with comparable results to piezoelectric sensors; however, this system needs a cloud and high computation effort for implementation [14]. Grimmelsman et. al evaluated the performance of an ADXL335 MEMS low-cost accelerometer with analog output for bridge vibration measurement. This accelerometer was tested under ten different harmonic excitations and on a typical highway bridge. The results were compared with an instrument-grade accelerometer that showed this developed system could identify the dynamic characteristics of a structure. They only considered the z-direction to compare the output, as this direction had the worst case in terms of noise density[15]. Komarizadehasl et al. proposed a new data acquisition system by combining five low-cost accelerometers. This proposed system shows high accuracy on low frequency and low acceleration amplitudes. Even though this CHEAP system has a very low error in frequency between 0.5-10 Hz, the fabrication of this sensor is not easy, depends on the computer, and has high noise density. It also measures the acceleration in one direction [16]. So, they upgraded their system and developed a new low-cost Adaptable Reliable Accelerometer (LARA) using Arduino technology that can be used in SHM. They tested the system on short-span Foot Bridge and compared the result with high-precision commercial sensor [17]. A low-cost acquisition system suggested by Meng et al. is made up of a Raspberry Pi 4 and an LSM9DS1 accelerometer. A commercial accelerometer (PCB 356B18) was used to test this system [18]. Previously, researchers developed a group of low-cost efficient wireless intelligent sensors (LEWIS) to provide a straightforward and affordable platform for SHM sensing and introducing new users to the world of sensors [19]–[21].

There are researchers, laboratory experiments, and ongoing topics which extensively investigate the use of low-cost sensors; however, the knowledge of sensor technology and the ability to build a ow-cost sensor without prior electrical or programming experience should be transferred to students and civil engineers. A practical demonstration or experiment with a link to contemporary issues can provide textbook material fresh significance and potential for students, as well as shed light on previously unconsidered career choices. LEWIS1 is an easy-to-use accelerometer and is considered the first iteration of LEWIS devices. Additionally, approachable methods are vital for educating new learners of STEM and civil engineering as it alleviates the intimidating aspects of these subjects. Once learners understand the exciting process of using sensors for projects such as robotics and earthquake monitoring, they may be open to continuing their education in STEM and civil engineering. Therefore, the authors believe that the LEWIS1 can be used not only for the engineering purposes, but also for educating and teaching the concept of technology with variety of application to students.

This paper summarizes the work of designing, developing, and testing a low-cost sensor by focusing on a new version of the LEWIS1 (called LEWIS1-β) that can be used for vibration measurement. A shaking table generated six different excitations and acceleration are recorded with LEWIS1 and LEWIS1-β sensors and commercial sensor. Moreover, it also summarizes the improvements and comparative testing of this low-cost sensor and a commercial sensor as a ground truth. The contents of this work enable other educators and researchers to use this platform in their institutions for multiple educational and outreach activities. Additionally, examples to date are shared to illustrate the feedback from the users.

## 2. LEWIS1

The Low-cost Efficient Wireless Intelligent Sensor, called LEWIS, is the first generation of low-cost sensing series that aims to resolve issues within structural health monitoring (SHM), assist in the decisions of infrastructure managers, and educate individuals on the concept of low-cost sensing. The simplistic architecture of the LEWIS sensor makes it to be low-cost and easy to fabricate even for the person without prior knowledge of electric and computer engineering. The components of this low-cost sensing platform that can measure the acceleration are described in the hardware section. Then, Arduino platform will be introduced that covers the software requirements for capturing acceleration data. Finally, the limitations and the ways for improvement will be discussed.

### 2.1. LEWIS1 Hardware

The LEWIS1 unit is composed of three main parts, microcontroller, sensing element, and development shield. The microcontroller is an ATmegaa328p-based called Arduino Uno R3 board with 14 digital I/O pins. A USB cable or an external power source can be used to power the Arduino board (such as an AD-to-DC adapter or a battery). A tri-axial MMA8451 digital accelerometer with a built-in 14-bit ADC serves as the sensing component [22]. Its usage spans a wide spectrum, from ±2g up to ±8g. The last part is the development shield which includes a breadboard that turns the Arduino and accelerometer into a circuit by carrying all Arduino pins to the upper layer. Figure 1 shows the eight main components of the LEWIS1.

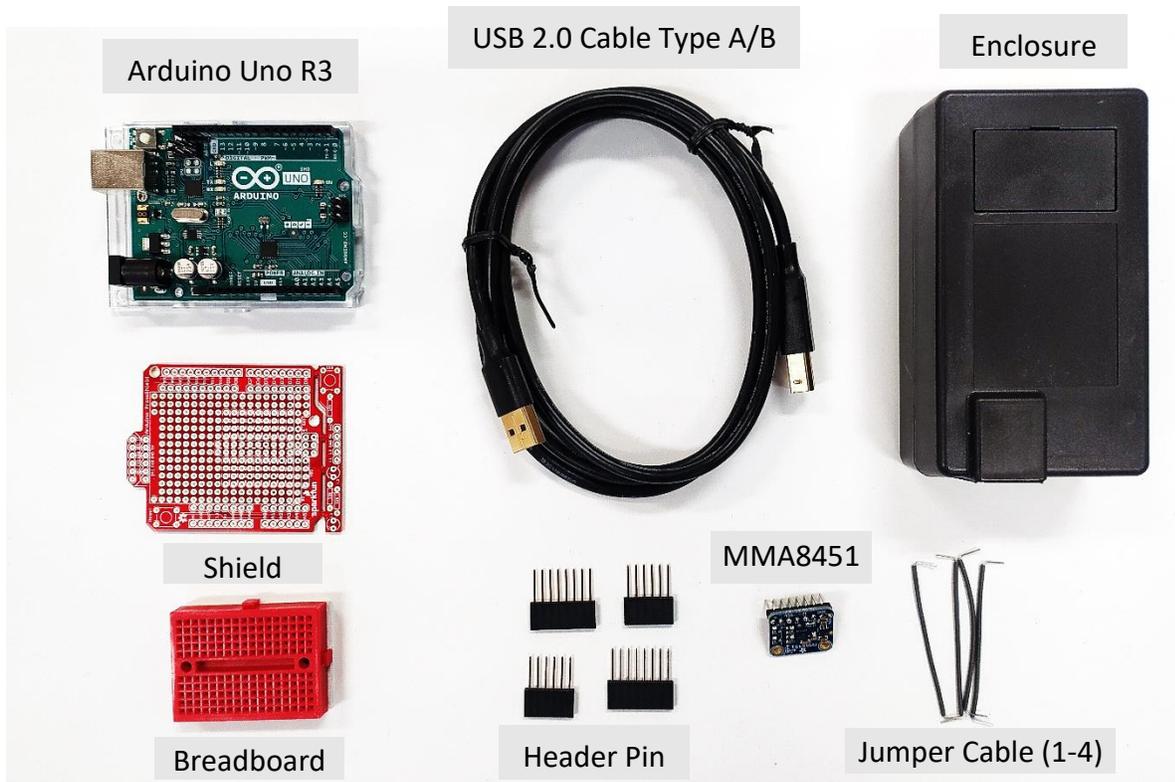

*Fig 1. LEWIS1 component*

Additionally, the researchers massed LEWIS1 building kits to further simplify the sensor assembly process. The building kit was comprised of the LEWIS1 components, a box casing to house the components, and access to an assembly and operations manual.

Table 1 shows the cost of the LEWIS1 components. Based on the table, the total cost of the LEWIS1 kit is less than $60 which demonstrates the cost-effectiveness of this sensor.

Table 1. The specification of the Arduino microcontroller

| Component | Price, $ |
| --- | --- |
| Arduino Uno R3 | $ 25.30 |
| MMA8451 Accelerometers | $ 7.95 |
| Breadboard and shield | $ 2.12 |
| USB 2.0 Cable Type A/B | $ 7.00 |
| Enclosure | $ 8.81 |
| **Total** | **$ 51.18** |

### 2.2. LEWIS1 Software

The Arduino integrated development environment (IDE) is a simple and easy-to-learn software for writing Arduino programs, also known as Arduino sketches. It is possible to write, compile, and deploy code on a board using IDE. This IDE supports a C/C++ dialect that employs unique code organization guidelines [23]. The researchers developed an Arduino sketch that allows the Arduino IDE software to read incoming sensor data. For this sketch, there are two libraries for this particular accelerometer that are required to be installed. Once the sensor USB cable is plugged into the computer, the provided Arduino sketch can be uploaded onto the Arduino board, which enables the user to read the acceleration in three different axes.

### 2.3. LEWIS1 Assembly Instruction

Building a sensor can be a challenging process, especially for individuals who are unfamiliar with the techniques and materials involved. However, with the right set of instructions and materials, it is possible to construct a sensor that can measure various parameters and provide valuable data.

One way to build a sensor is to follow a step-by-step guide or instructions that are provided in a figure as presented in Figure 2.

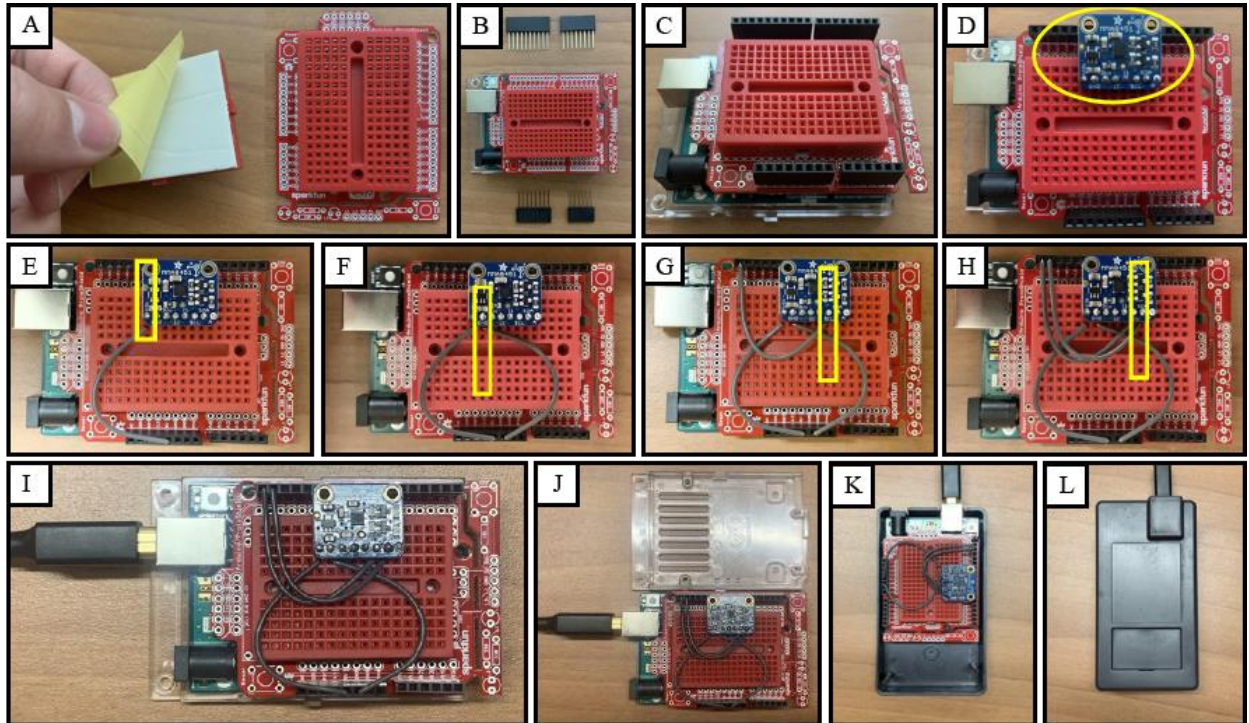

*Fig 2. LEWIS1 assembly steps*

### 2.4. Limitations and Improvement

The LEWIS1 capabilities bring new opportunities for students and educators to perform experiments with their own built sensor without having extensive knowledge of electrical components. The researchers trained educators and new learners to build LEWIS1s. The users provided suggestions to resolve difficulties experienced during the assembly. To enhance the potential capacity of the LEWIS1, a new iteration of LEWIS1 can be developed to address some of the challenges of the present version and improve it. The LEWIS1 needs soldering the accelerometer to the header pin, which is hard for students and requires soldering tools. The assembly of the sensor can be easier by replacing components that require soldering. Currently, four wires need to be connected to the header pins, which can be replaced by one wire with four outputs (1 to 4). Also, other degrees of freedom, such as measuring angular velocity and temperature, could be added to the capability of LEWIS1. Additionally, the software has the ability to be programmed in a way that makes it much easier for the user to install, import, and execute the code. Hence, the next section will introduce and summarize the use of a new accelerometer that could be an alternative to the LEWIS1 accelerometer.

### 3. LEWIS1 Upgrade (LEWIS1-β)

The LEWIS1 is capable of measuring tri-axis acceleration with a low-cost, portable, small-size platform. Different accelerometers can be replaced to build a new platform for measuring acceleration. The authors propose a new LEWIS1 called LEWIS1-β that has advantages over the previous sensor. The new platform architecture is improved by replacing the accelerometer, providing simpler assemblage, and using simpler software. One of the most important differences between these two sensors is that the LEWIS1-β not only measures the tri-acceleration but also has the capability of measuring the temperature and the angular velocity since there is a built-in gyroscope in the sensor. This will broaden the usage of this new sensor and bring lots of potential applications for new projects. Additionally, the LEWIS1-β is much simpler to fabricate as it has fewer components. In the following, further information about the software, hardware, and user-teaching interface of the new sensor will be provided.

### 3.1. LEWIS1-β Hardware

The newly developed LEWIS1 is comprised of two primary components, a microcontroller, and a sensing element. The microcontroller board is an Arduino Uno SMD R3 based on the ATmega328p single-chip microcontroller, which has 14 digital I/O pins that are utilized in the new sensor platform. The new sensing element is the LSM6DSOX, which is a 6-degree of freedom (DOF) inertial measurement unit (IMU) that includes an accelerometer and gyroscope [24]. The accelerometer has a decent range of $\pm2/\pm4/\pm8/\pm16$ g at a 1.6 Hz to 6.7 kHz update rate. Figure 3 shows all the required components for the assembly of LEWIS1-β.

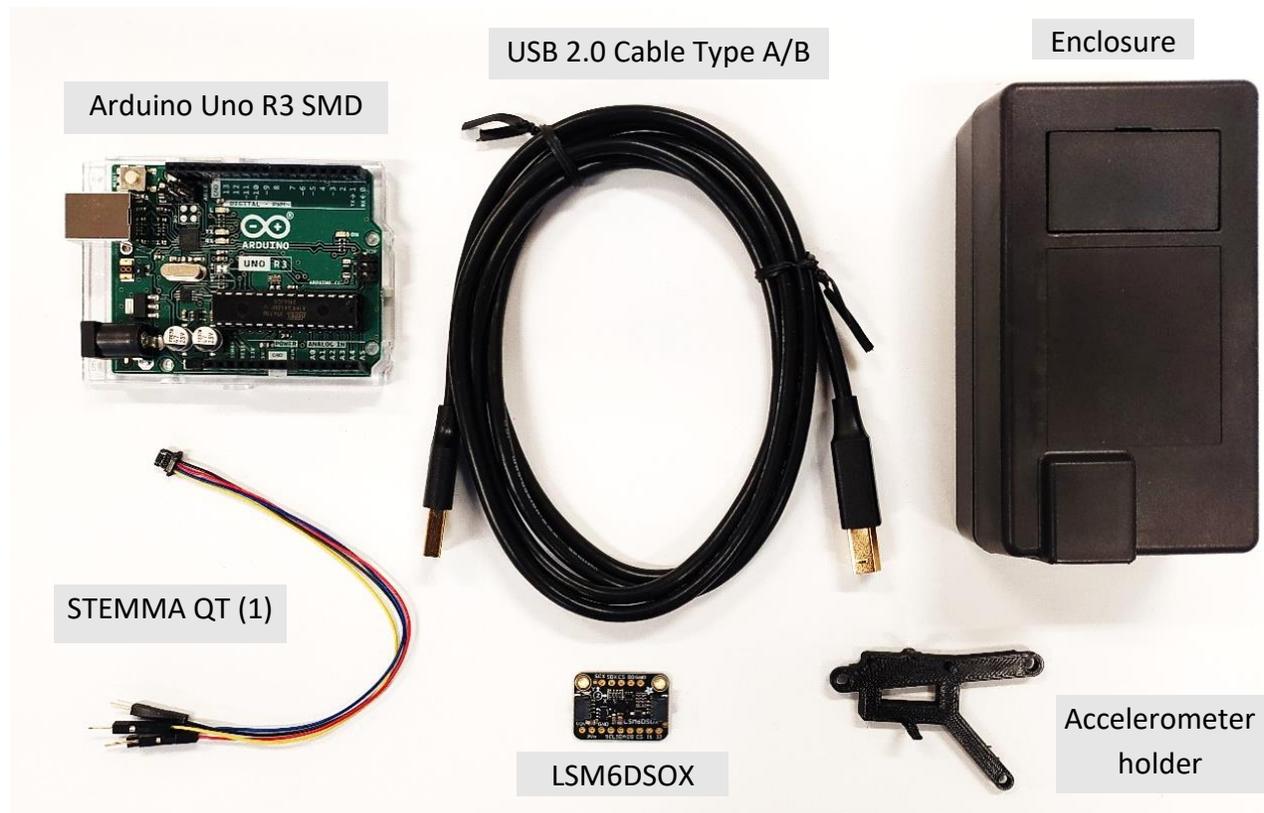

*Fig 3. LEWIS1-β component*

Table 2 shows the cost of the LEWIS1 components. This table shows that even though this sensor has more capability compared to the LEWIS1, however; the total cost of a kit containing all the components is less than $60.

Table 2. Cost of LEWIS1-β components

| Parameters | Details |
|---|---|
| Arduino Uno R3 SMD | $ 24.10 |
| LSM6DSOX Accelerometers | $ 11.95 |
| STEMMA QT / Qwiic JST SH 4-pin to Premium Male Headers Cable | $ 0.95 |
| USB 2.0 Cable Type A/B | $ 7.00 |
| Enclosure | $ 8.81 |
| **Total** | **$ 52.81** |

The Arduino microcontroller is used to create computers that can interact with their surroundings, collect information via sensors, and operate objects like lights and motors accordingly. Researchers chose Arduino due to its adaptability and simplicity in linking with additional sensors like accelerometers, gyroscopes, and magnetometers, as well as its widespread market availability for the end user. Additionally, Arduino features a built-in development open-source platform that supports the C programming language. The term "Open Source" denotes that all the board's resources, including the design and CAD files, are free and available to everyone [25]. That enables users to freely tune the microcontroller's performance through a USB connection. This implies that anyone can change it to suit their needs. With Arduino, both professionals and students may easily and affordably build microcontroller computers that can communicate with their surroundings.

To assemble all the components of LEWIS1-β together and enclose them in the box, a holder plate is designed to further simplify the assembly process. The holder is a small plastic plate that securely holds the sensing element. The holder plate is designed to consider the lack of a breadboard in the LEWIS1- β sensor. The plate is composed of a central rectangular part and three connected arms. The central part is the housing location of the sensing element. This part utilizes two snap clips along with two small protruding cylinders to effortlessly secure the sensing element to the plate. Each arm contains a 3.5 mm hole at the very end which are used to affix the plate to the new LEWIS1-β enclosure. Figure 4 demonstrates the top and isometric view of the designed plate.

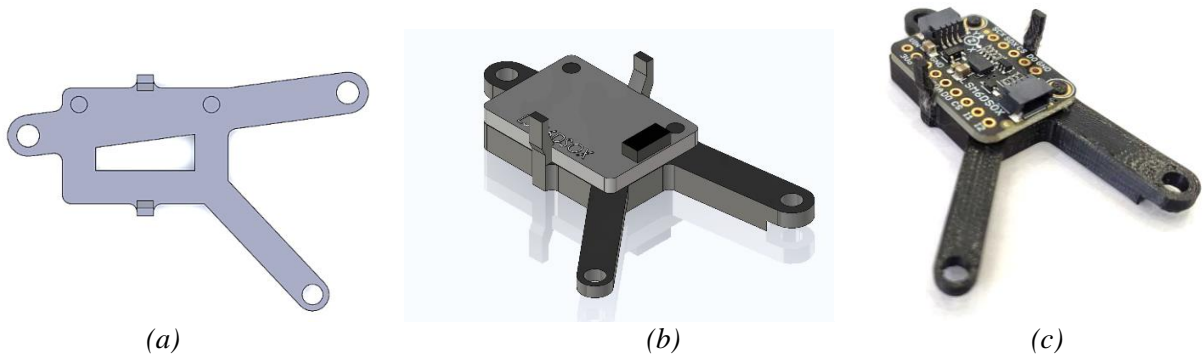

*(a)* *(b)* *(c)*

*Fig 4. LEWIS1-β 3D printed holder (a) Top view, (b) Isometric view, (c) real photo*

### 3.2. LEWIS1-β Software

The LEWIS1-β sensor is also operated with the Arduino IDE. The LEWIS1- β sensor uses an Arduino sketch that is different than the previous version as the new sensing element requires a different method of processing data. Three libraries are required to be installed to read and plot the data. Once the user installed the three libraries, they will be able to read, visualize, and record the data.

## 3.3. LEWIS1-β Assembly Instruction

Figure 5 shows all the steps required for a person to follow being able to build the LEWIS1-β sensor. Comparing the steps with the LEWIS1 it shows that the number of steps and complexity of assembly is reduced for the user. Hence, the user would be able to do the process in shorter time and with less errors.

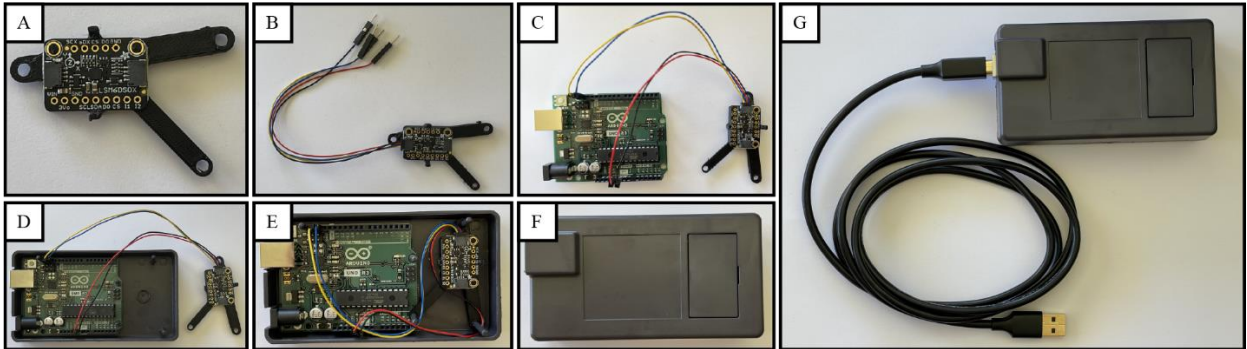

*Fig 5. LEWIS1-β assembly steps*

## 4. LEWIS1 and LEWIS1-β Comparison

The assemblage of the new sensor has some differences from the old one. First, the LEWIS1-β sensor does not need a breadboard and shield since the accelerometer is directly mounted on the 3D-printed plate. Second, without a breadboard, the soldering of header pins to the accelerometer would not be required. Third, instead of four single wires, a 4-pin male header jumper wires will be used, which reduces the time and complexity of wiring. Figure 6 shows the assembly of the two LEWIS sensors.

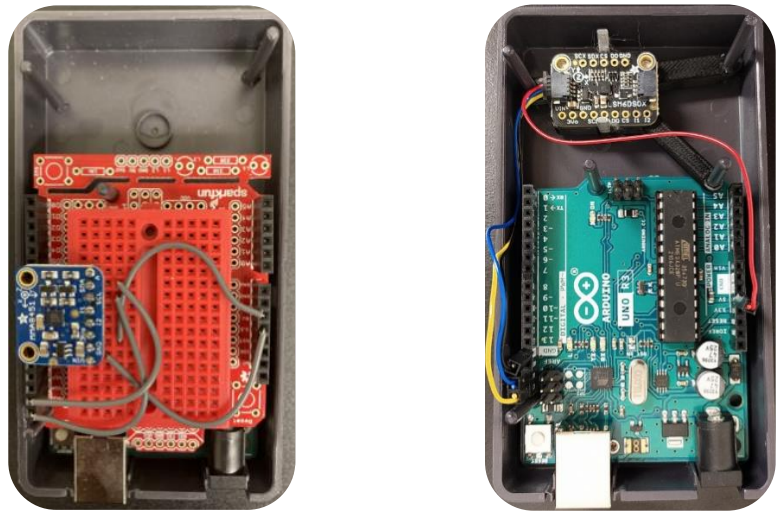

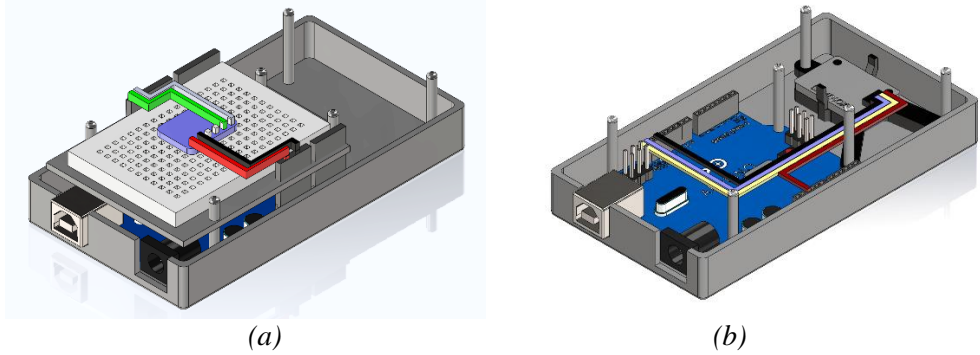

*(a)*            *(b)*
*Fig 6. (a) LEWIS1, and (b) LEWIS1-β plan view comparison*

After comparing the components required for the LEWIS1 and LEWIS1-β sensors, it's clear that the newer version has advantages in terms of component count. The LEWIS1-β sensor requires only nine components, while the older LEWIS1 sensor requires 18 components. This reduction in component count can lead to several benefits, such as lower manufacturing costs and time, improved reliability, and increased ease of assembly. Additionally, the fact that the LEWIS1-β sensor requires only one wire connection instead of four may simplify the wiring process and reduce the risk of errors. Table 3 shows the number of component comparison.

Table 3. LEWIS1 and LEWIS1-β required number of components

| Components | LEWIS1 | LEWIS1-β |
|---|---|---|
| **Arduino board** | 1 | 1 |
| **Accelerometer** | 1 | 1 |
| **wire connection** | 4 to 4 | 1 to 4 |
| **Breadboard** | 1 | 0 |
| **Shield** | 1 | 0 |
| **Header pin** | 4 | 0 |
| **USB cable** | 1 | 1 |
| **Enclosure** | 1 | 1 |
| **Total** | 18 | 9 |

Figure 7 shows the comparison regarding the number of components needed to be built in order to assemble one sensor of each type. As it can be seen in the figure, LEWIS1 required is more complex, it also needs four different wire to be connected to breadboard using 4 separate header pin. While LEWIS1-β uses a 4 to 1 cable which reduce the wiring complexity and it does not need header pin as they are mounted on the Arduino themselves.

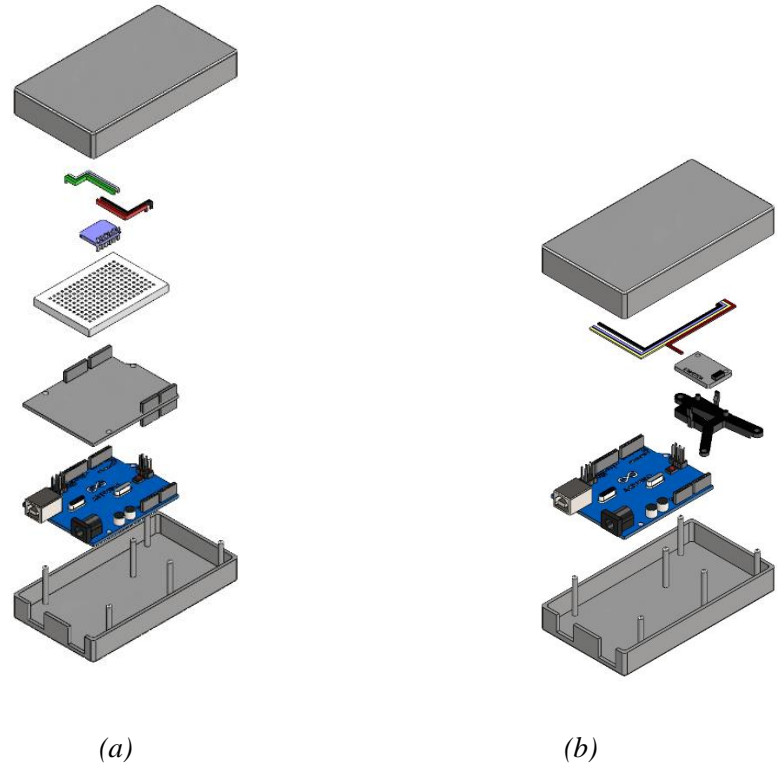

*(a)                    (b)*
*Fig 7. (a) LEWIS1, and (b) LEWIS1-β number of components*

Table 4 summarizing the properties of two version of LEWIS sensors. The LEWIS1-β has been improved in many aspects including the reduction of number of components which makes the fabrication simpler and faster while maintaining almost the same cost. It has wider range of acceleration usage and can be also used for calculating angular velocity. One factor that makes the fabrication of LEWIS1-β harder than the LEWIS1 is that it has a component that needs to be 3D printed for holding the accelerometer while the LEWIS1 only requires soldering. Generally, both accelerometers are easy and quick to fabricate specially for students without any background of electrical and computer engineering, and they provide the accuracy comparable to commercial sensor.

Table 4. LEWIS1 and LEWIS1-β comparison

| Properties | LEWIS1 | LEWIS1-β | Improved? |
|---|---|---|---|
| Accelerometer | MMA8451 | LSM6DOX | - |
| No. of components | 14 | 6 | ✓ |
| ADC | 14 bit | 16 bit | ✓ |
| Accelerometer range | ±2g/±4g/±8g | ±2g/±4g/±8g/±16g | ✓ |
| Gyroscope range | - | ±125/±250/±500/±1000/±2000 dps | ✓ |
| Fabrication time | 20 mins | 10 mins | ✓ |
| Cost | $51.18 | $52.81 | ✓ |
| Special requirement | Soldering the header pin to accelerometer | 3D printing the holder | - |
| Application | 3DOF Accelerometer | 3DOF Accelerometer, 3DOF Gyroscope, Temperature | ✓ |

## 5. Validation and characterization

Researchers carried out an experiment to demonstrate the capability of the LEWIS1 platform compared with the LEWIS1. A commercial reference accelerometer was used as ground truth to prove the accuracy and effectiveness of the proposed measuring platform. The actions were done to validate the suggested sensing platform are described in this section. The z axis in general is the main measurement axis which the data will be collected using the LEWIS1 platform once it is attached. The excitation inputs and the performance evaluation standards that were used to test and evaluate the suggested sensing platform are outlined. The final section examines the data gathered from the experiments and makes comparisons between the suggested system, LEWIS1, and the commercial sensor measurement.

### 5.1. Experiment Setup Description

In this study, the LEWIS1 and new suggested LEWIS1-β sensing platform are compared to a uni-axis #353B33 ICP accelerometer (DC type) from Piezotronics capacitive sensor. Two sensors are mounted on a APS 113 Shaker, which can provide up to 133N peak sine force with up to 158 mm stroke. Both LEWIS sensors are coupled to a VibPilot DAQ system which is an eight channel data acquisition system produced by M+P International. The VibPilot has a 24-bit A/D converter with antialiasing sampling rates up to 102.4 kHz.

Each LEWIS sensor should be connected to a computer to collect the data by running the IDE application. The PCB sensor needs to be connected to the DAQ system to read and record the data. Therefore, one channel of the VibPilot is allocated to the PCB sensor as a response, and the smart shaker is also connected to one channel as an excitation input. Figure 8 shows the connection of the entire system to the DAQ platform.

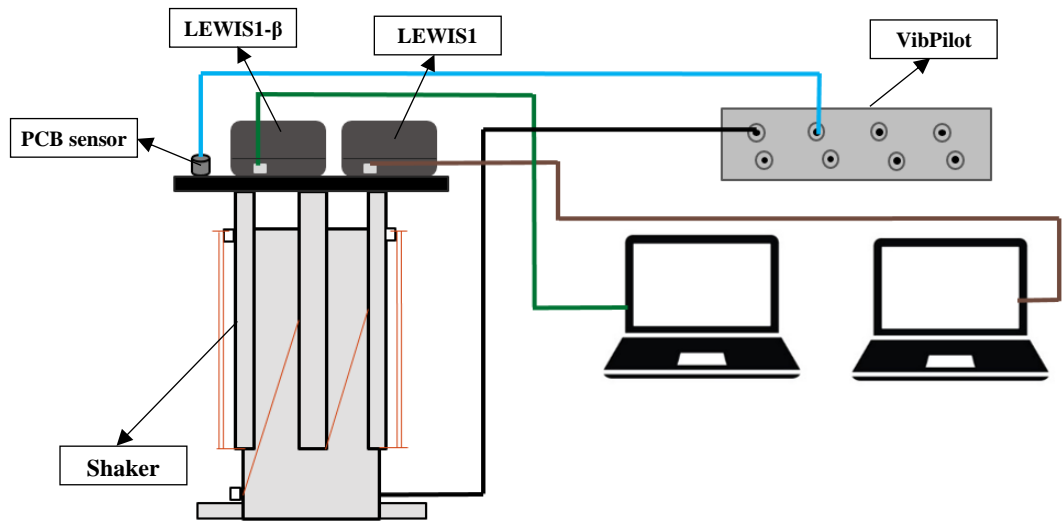

*Fig 8. Experiment data collection platform*

To be able to assemble all the three sensors together and attach them to the smart shaker, a plate was designed and 3D printed. That plate and the clamps keep the entire system together. Therefore, the same excitations will be applied to all three sensors simultaneously. The collected data will be saved and further analyzed with MATLAB software. A picture of the experimental setup, which includes the shake table, the two LEWIS sensors (MMA8451 and LSM6DS), and the commercial PCB accelerometer is shown in Figure 9.

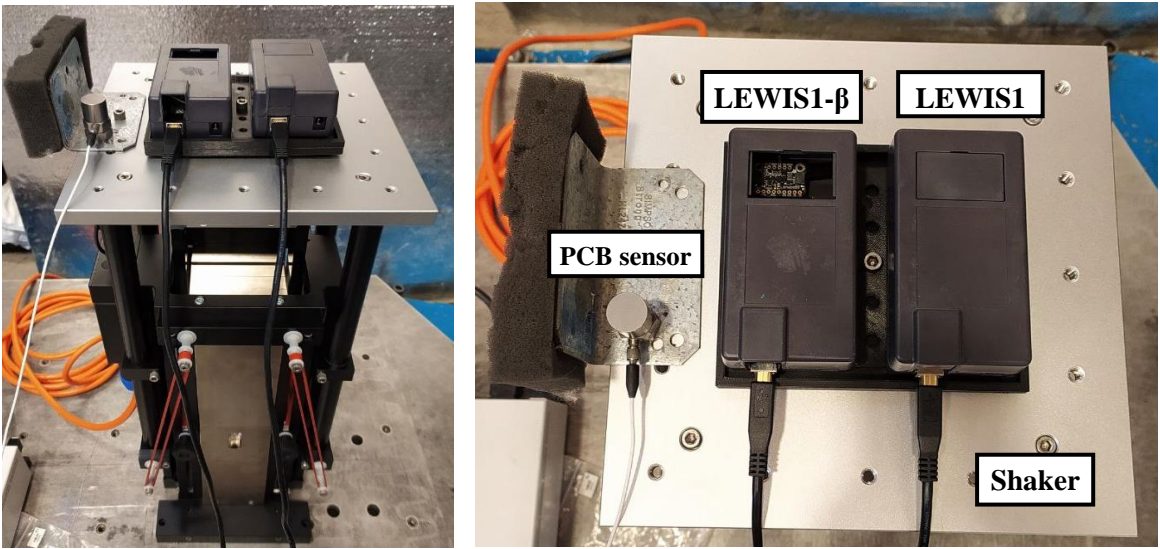
*Fig 9. Experiment setup*

### 5.2. Excitation Input

The two sensors and the PCB accelerometer were mounted on the smart shaker that was connected to VibPilot to excite the three sensors at the same time. A total of six different excitations were selected as an input to the system: (1) 1 Hz Sinusoidal excitation, (2) 3 Hz Sinusoidal excitation, (3) 5 Hz Sinusoidal excitation, (4) 10 Hz Sinusoidal excitation, (5) Sine sweep excitation with the frequency range 0-10 Hz, and (6) the Band-limited White Noise (BLWN) with the cap of 10 Hz. For all excitation scenarios, the readings from LEWIS accelerometers and PCB sensor were captured at 100 Hz, and 1024 Hz, respectively. For each stimulation case, the measurement data were gathered for 16 seconds. Table 5 shows the list of excitations.

Table 5. Excitation specification

| Type of excitation | Frequency (Hz) |
|---|---|
| Sinusoidal | 1 |
| | 3 |
| | 5 |
| | 10 |
| Sine sweep | 0-10 |
| Band-limited white noise | 10 |

### 5.3. Result and Analysis

Researchers applied six different excitation to the shake table to excite the LEWIS1, LEWIS1-β, and PCB sensor mounted on the plate, simultaneously. LEWIS sensors captured data at 100 Hz, and data were logged in the computer connected to each of them. The VibPilot generated a response and sent the excitation to the smart shaker. The response of the PCB accelerometer with a 1024 Hz sampling rate was recorded by VibPilot. It is important to keep in mind that this test carried out to obtain the findings in this section had to be manually initiated and stopped for the LEWIS sensors at or around the same time, which led to some minor but unavoidable discrepancies in the time vectors. To remove the effects of sudden movement and noises at the start and the end of the whole excitation time, the first and last two seconds of the response was deleted. Hence, the experiment was analyzed for 12 seconds. The time-domain result of the two sensor's responses and the PCB for band-limited white noise is shown in Figure 10. While multiple experiments were conducted with various excitations using mentioned sensors and shaker, this paper focuses on presenting one experiment as a representative example of the study's key results as the main purpose of this paper is to represent the performance of this sensor in educating different people having different level and knowledge of electrical and computer background.

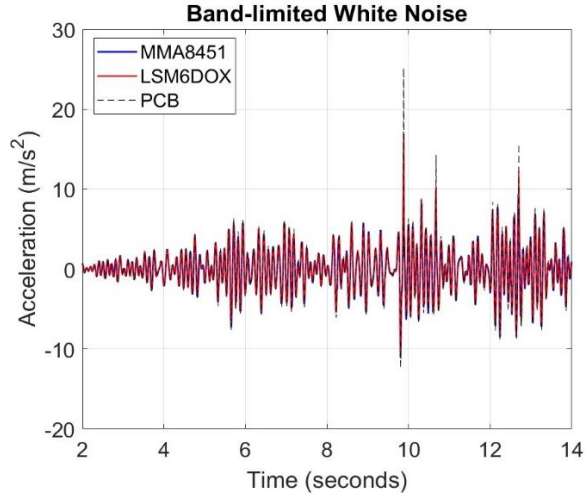

*Fig 10. Time-domain response of band-limited white noise.*

As the sampling rate of the LEWIS sensors and the reference accelerometer are different, for analysis and fair comparison, the data should be resampled at the LEWIS sampling rate which is 100 Hz. To be able to compare the performance quality of the two sensors with the PCB as ground truth, two indices are calculated. The root-mean-square error (RMSE) is calculated for each LEWIS and reference accelerometer which computes how far the measurements of new signal from the measured PCB sensor as a true values are. The equation given below is used to calculate RMSE.

$$RMSE = \sqrt{\frac{\sum_{i=1}^{n}(acc_{PCB}-acc_{LEWIS})^2}{n}} \qquad (1)$$

Where,
$acc_{PCB}$ and $acc_{LEWIS}$ are the acceleration of PCB and LEWIS accelerometers, respectively.
$n$ represents the number of data points.
The percentage difference is calculated based on the following equation:

$$Percentage\ difference\ (\%) = \frac{|acc_{PCB}-acc_{LEWIS}|}{(acc_{PCB}-acc_{LEWIS})/2} * 100 \qquad (2)$$

The second indexes for comparing two signals is signal-to-noise ratio (SNR). SNR represents which sensor has the most accurate response. The SNR can be calculated using equation (3).

$$SNR = 10\ log\ \left(\frac{Signal\ power}{Noise\ power}\right) \qquad (3)$$

Table 6. SNR of MMA8451, LSM6DOX, and PCB

| Type of excitation | Index | Frequency (Hz) | MMA8451 | LSM6DOX | PCB | Difference (%) |
|---|---|---|---|---|---|---|
| Band-limited white noise | RMSE | 10 | 6.0501 | 6.2694 | - | 3.5602 |
| | SNR | 10 | 12.4792 | 14.6635 | 27.1961 | 16.0949 |

Based on the Table 6, the RMSE of LEWIS1 is higher than LEWIS1-β which indicates that the performance of the LEWIS1-β is better than the LEWIS1. Table 6 also shows the signal-to-noise ratio of two LEWIS sensors and PCB. Larger SNR values indicate that the signal strength is stronger relative to the noise levels, allowing for higher data rates and fewer retransmissions, which together improve throughput. Based on Table 6, the PCB sensor had the highest SNR values, whereas the LEWIS1 had the lowest SNR values. The LEWIS1-β sensor, however, had SNR values that were on average 23% higher than those of the LEWIS1. This means that the LEWIS1-β sensor produced the responses with the lower amount of noise rather than LEWIS1. The SNR can be raised by reducing noise. External (electromagnetic), conducted, and intrinsic noise are some of the several sources of noise [26]. While conducted and external sounds can be eliminated or decreased with the right setup and circuit designs, intrinsic noise cannot be eliminated but can be minimized with the right heater and thermal sensor element design.

## 6. Integration of Sensor Technology into Civil Engineering Education Curriculum

The integration of sensor technology into civil engineering education curriculum is an effective way to enhance students' learning experiences. By incorporating sensors into the curriculum, students can gain practical skills in data collection and analysis, which are essential for success in civil engineering. The use of sensors can provide valuable information on structural performance and safety. This data can aid in the design and maintenance of infrastructure, contributing to the development of safe and sustainable civil engineering projects. Moreover, the integration of sensor technology can help students stay up-to-date with the latest advancements and innovations in the field, further enhancing their career prospects. As such, the inclusion of sensor technology into civil engineering education curriculum can have a significant impact on students' career readiness and the development of safe, sustainable infrastructure.

The lack of widespread use of sensors in education can be attributed to several factors. Firstly, traditional methods of data collection and analysis often require extensive manual labor and are time-consuming, making it difficult for educators to incorporate them into the curriculum. Additionally, sophisticated sensor technology has historically been expensive and often out of reach for many educational institutions. However, the emergence of low-cost sensors has significantly lowered the barriers to entry for educators and students, making this technology more accessible and easier to use.

The benefits of incorporating sensor technology into education are numerous. By using sensors, students can gain hands-on experience in data collection and analysis, which is an essential skill in many fields, including civil engineering. Additionally, the use of sensors can provide real-time data that can aid in problem-solving and decision-making. For instance, in civil engineering, sensors can help identify structural problems or areas that require maintenance before they become critical, potentially saving lives and reducing costs. Moreover, the use of sensor technology can help students develop critical thinking skills as they interpret and analyze data, promoting a deeper understanding of concepts and theories.

## 7. User Outreach and Feedback

A mobile, distributable field learning and training environment that can serve as a model for other educational institutes and programs working on the same topic has been developed. Sensors are one of the tools that can collect experimental and field data measurements to quantify technical and scientific information. There are many areas where sensors can be used. Therefore, the authors used the LEWIS sensor to engage students involved in outreach activities and educational program to sense and measure the vibration of their surrounding area. The platform that is being utilized to teach the children how to build and use the sensor is slightly different in terms of software development. As the students use Chromebooks instead of regular laptops, the authors ensured that the new developed platform is compatible with Chromebooks. The difference between the software on these two systems is that the Chromebook uses the "Arduino Create for Education" application and a Chrome extension for plotting data. Otherwise, the performance of both systems is similar.

The advantage of the LEWIS sensor is that it is simple for kids to fabricate. Without any prior knowledge of programming, they can execute the code to see the results of environment sensing measurement. The advantage of being low-cost makes this sensor affordable for all students. Therefore, the research team used this sensor as an educational tool for K-12th students, civil engineers, or people in any professions to engage them in different activities. The researchers at the University of New Mexico's Smart Management of Infrastructure Laboratory (SMILab) developed sensor classes and workshops for students, engineers, and educators since 2015. The purpose of this activity is to assemble the LEWIS sensor and improve hands-on technical skills. Then, they employ that sensor in real-world applications. This will boost the person's adoption of research and increase their creativity as they can think of how this sensor can be used in different areas. Additionally, the K-12th students and learners of different ages have the opportunity to experience exposure to industry and also think about integrating new technologies with existing infrastructure.

Table 7 shows the number of people with different ages and professions that attended to sensor fabrication workshops since 2015 and also the number of sensors that has been built. Based on Figure 11 the number of sensors and people participating in sensor workshop has ascending trend except for year 2019, 2020, and 2021. In these three mentioned years the UNM team was not able to hold the in-person sensor workshop due to COVID; however, there were online classes for students where they had the sensor kit in their home and the instructor taught them the fabrication process.

Table 7. Number of people and built LEWIS sensor from 2015 to 2022

| Year | No. of sensors | No. of people |
|------|----------------|---------------|
| 2015 | 1 | 1 |
| 2016 | 6 | 6 |
| 2017 | 22 | 52 |
| 2018 | 43 | 57 |
| 2019 | 31 | 35 |
| 2020 | 7 | 7 |
| 2021 | 27 | 29 |
| 2022 | 66 | 135 |
| **Total** | **203** | **322** |

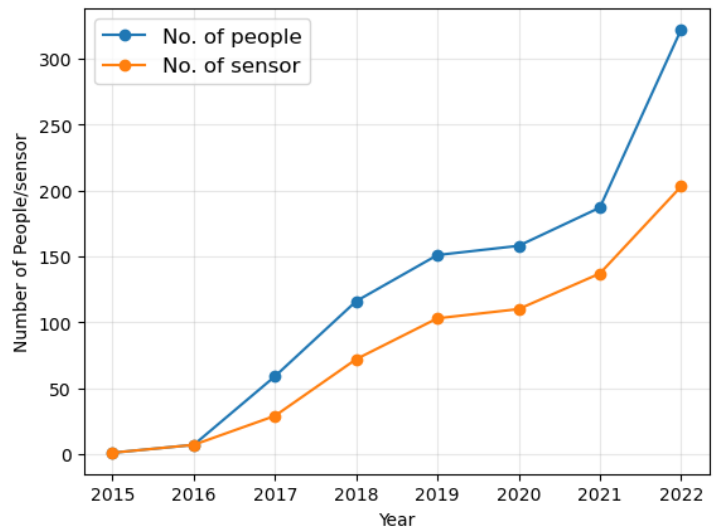

Fig 11. Cumulative graph of number of people and sensor (2015-2022)

Figure 12 demonstrates the percentage of different groups including the elementary school, middle school, high school, and college/university students, and also the people different professions, who attended the sensor building workshops. Half of the attendees were among the college/university students with different major. High school, professionals, middle school and elementary schools were participated 21%, 15%, 10%, and 4% respectively.

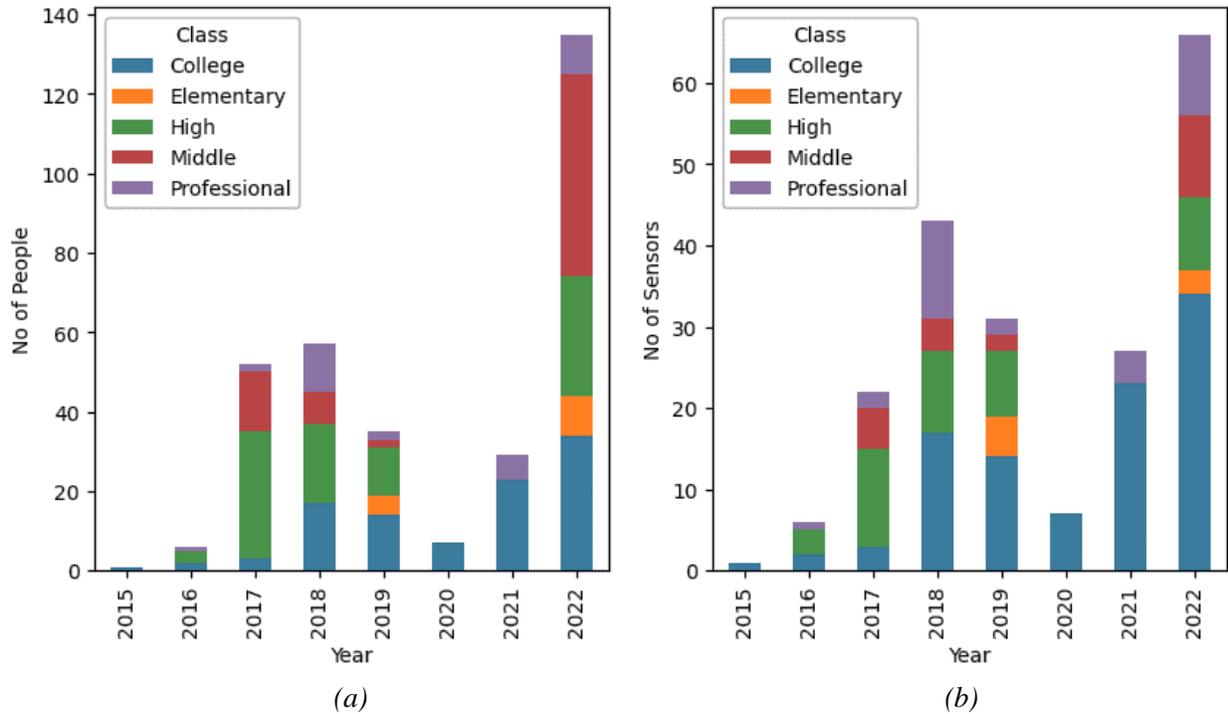

*Fig 12. Participation of different class of people from 2015-2022 (a) Number of people, and (b) Number of sensor*

The students built the LEWIS sensor and tested it in different places such as Albuquerque Rail runner, Sandia Peak Tramway, and the University of New Mexico Duck Pond Bridge. Figure 13 shows the LEWIS workshop and an experiment that has been done in Albuquerque Rail runner.

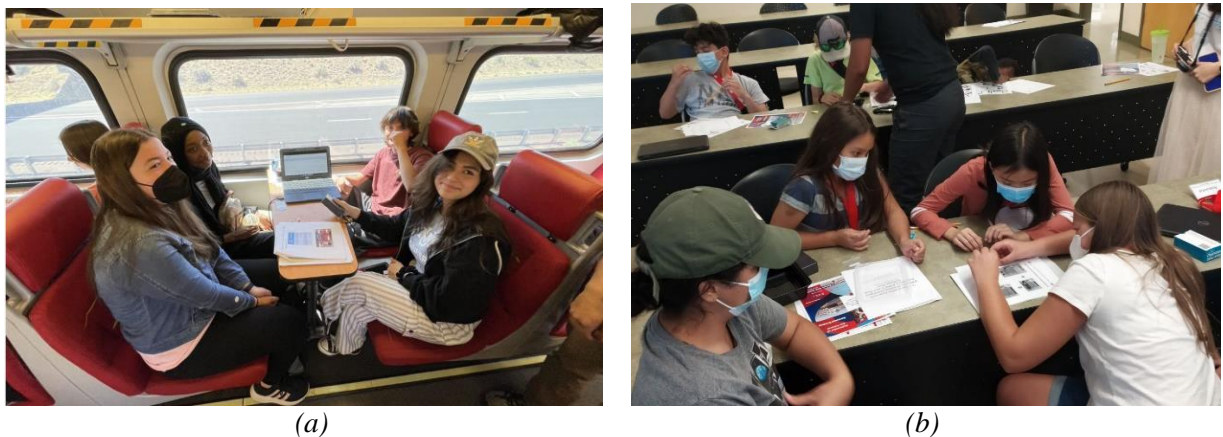

*Fig 13. Outreach activities: (a) LEWIS on Railrunner, (b) LEWIS workshop*

Exploring new Arduino technologies and building sensors has fostered human development from the earliest phases of education. The findings demonstrated that after taking part in the design and production of low-cost sensors, students at all levels developed an interest in infrastructure maintenance methods.

## 8. Conclusions

This paper introduced a new low-cost sensor (LEWIS1-β) that has the capability of gathering experimental and field data measurements. This new sensor is the upgraded version of the LEWIS sensor that had been used before as a tri-accelerometer. The developed sensor can be used to measure the tri-axis acceleration, rotation, and temperature with a low-cost, portable, small-size, and user-friendly system. LEWIS1-β offers the advantage of maintaining a simple design, allowing researchers with limited knowledge of sensor technology to quickly fabricate and use the sensor. The accuracy of these sensors in collecting data was validated using a commercial PCB sensor. All three sensors were mounted on the smart shaker and three different excitation types, sinusoidal, sine sweep, and band-limited white noise were applied to them. The RMSE and SNR, as two indicators of differences between each sensor and PCB, were calculated to illustrate that the LEWIS1-β has comparable performance compared to the expensive commercial accelerometer. It also shows better accuracy than the LEWIS1 with almost the same cost. The results of this study show that low-cost sensor can potentially be used for structure response measurement as it has comparable accuracy and performance to the commercial sensor. It gives the ability of fabricating and using the sensors to the civil engineers who usually do not build the sensor for their profession's purposes.

The authors also used this sensor to implement different outreach activities aimed at training and educational purposes. The LEWIS1-β is quite simple to fabricate, easy to be programmed, and can be used for different applications. Therefore, the researcher used the LEWIS sensor to engage K-12[th] students, civil engineers, and people with different level of knowledge in outreach projects and sensor building workshop so they could feel and quantify the vibrations in their environment and be able to use the sensor into their day-to-day life.

**Acknowledgments**

The authors of this paper acknowledge the support from the Department of Civil, Construction, and Environmental Engineering of the University of New Mexico; and the Center of Advanced Research and Computing. The authors would also like to appreciate the continuous help and assistance from Eric Robbins, Odey Yousef, and Morgan Merill.

**References:**

[1] "The Vision for Civil Engineering in 2025," p. 108.

[2] N. Hoult *et al.*, "Wireless sensor networks: creating 'smart infrastructure,'" *Proceedings of the Institution of Civil Engineers - Civil Engineering*, vol. 162, no. 3, pp. 136–143, Aug. 2009, doi: 10.1680/cien.2009.162.3.136.

[3] S. M. Barnett and S. J. Ceci, "When and where do we apply what we learn?: A taxonomy for far transfer.," *Psychological Bulletin*, vol. 128, no. 4, pp. 612–637, 2002, doi: 10.1037/0033-2909.128.4.612.

[4] B. A. Sundaram, K. Ravisankar, R. Senthil, and S. Parivallal, "Wireless sensors for structural health monitoring and damage detection techniques," *Current Science*, vol. 104, no. 11, pp. 1496–1505, 2013.

[5] C. Z. Li, Z. Guo, D. Su, B. Xiao, and V. W. Y. Tam, "The Application of Advanced Information Technologies in Civil Infrastructure Construction and Maintenance," *Sustainability*, vol. 14, no. 13, Art. no. 13, Jan. 2022, doi: 10.3390/su14137761.

[6] S. Komarizadehasl, B. Mobaraki, H. Ma, J.-A. Lozano-Galant, and J. Turmo, "Low-Cost Sensors Accuracy Study and Enhancement Strategy," *Applied Sciences*, vol. 12, no. 6, Art. no. 6, Jan. 2022, doi: 10.3390/app12063186.

[7] T. Peng, M. Nogal, J. R. Casas, and J. Turmo, "Role of Sensors in Error Propagation with the Dynamic Constrained Observability Method," *Sensors (Basel)*, vol. 21, no. 9, p. 2918, Apr. 2021, doi: 10.3390/s21092918.

[8] Y. Zhuang, W. Chen, T. Jin, B. Chen, H. Zhang, and W. Zhang, "A Review of Computer Vision-Based Structural Deformation Monitoring in Field Environments," *Sensors (Basel)*, vol. 22, no. 10, p. 3789, May 2022, doi: 10.3390/s22103789.

[9] E. Z. Berglund *et al.*, "Smart Infrastructure: A Vision for the Role of the Civil Engineering Profession in Smart Cities," *Journal of Infrastructure Systems*, vol. 26, no. 2, p. 03120001, Jun. 2020, doi: 10.1061/(ASCE)IS.1943-555X.0000549.

[10] A. S. Rao *et al.*, "Real-time monitoring of construction sites: Sensors, methods, and applications," *Automation in Construction*, vol. 136, p. 104099, Apr. 2022, doi: 10.1016/j.autcon.2021.104099.

[11] J. P. Lynch, "An overview of wireless structural health monitoring for civil structures," *Philosophical Transactions of the Royal Society A: Mathematical, Physical and Engineering Sciences*, vol. 365, no. 1851, pp. 345–372, Feb. 2007, doi: 10.1098/rsta.2006.1932.

[12] A. B. Noel, A. Abdaoui, T. Elfouly, M. H. Ahmed, A. Badawy, and M. S. Shehata, "Structural Health Monitoring Using Wireless Sensor Networks: A Comprehensive Survey," *IEEE Communications Surveys & Tutorials*, vol. 19, no. 3, pp. 1403–1423, 2017, doi: 10.1109/COMST.2017.2691551.

[13] "(PDF) Title: Validation of an Integrated Network System for Real-Time Wireless Monitoring of Civil Structures." https://www.researchgate.net/publication/244411516_Title_Validation_of_an_Integrated_Network_System_for_Real-Time_Wireless_Monitoring_of_Civil_Structures (accessed Oct. 28, 2022).

[14] A. Girolami, F. Zonzini, L. De Marchi, D. Brunelli, and L. Benini, "Modal Analysis of Structures with Low-cost Embedded Systems," in *2018 IEEE International Symposium on Circuits and Systems (ISCAS)*, May 2018, pp. 1–4. doi: 10.1109/ISCAS.2018.8351705.

[15] K. A. Grimmelsman and N. Zolghadri, "Experimental Evaluation of Low-Cost Accelerometers for Dynamic Characterization of Bridges," in *Dynamics of Civil Structures, Volume 2*, Cham, 2020, pp. 145–152. doi: 10.1007/978-3-030-12115-0_19.

[16] S. Komarizadehasl, B. Mobaraki, H. Ma, J.-A. Lozano-Galant, and J. Turmo, "Development of a Low-Cost System for the Accurate Measurement of Structural Vibrations," *Sensors*, vol. 21, no. 18, Art. no. 18, Jan. 2021, doi: 10.3390/s21186191.

[17] S. Komarizadehasl, F. Lozano, J. Lozano-Galant, G. Ramos, and J. Turmo, "Low-Cost Wireless Structural Health Monitoring of Bridges," *Sensors (Basel, Switzerland)*, vol. 22, Jul. 2022, doi: 10.3390/s22155725.

[18] Q. Meng and S. Zhu, "Developing IoT Sensing System for Construction-Induced Vibration Monitoring and Impact Assessment," *Sensors*, vol. 20, no. 21, Art. no. 21, Jan. 2020, doi: 10.3390/s20216120.

[19] A. I. Ozdagli, B. Liu, and F. Moreu, "Low-cost, efficient wireless intelligent sensors (LEWIS) measuring real-time reference-free dynamic displacements," *Mechanical Systems and Signal Processing*, vol. 107, pp. 343–356, Jul. 2018, doi: 10.1016/j.ymssp.2018.01.034.

[20] M. Aguero, A. Ozdagli, and F. Moreu, "Measuring Reference-Free Total Displacements of Piles and Columns Using Low-Cost, Battery-Powered, Efficient Wireless Intelligent Sensors (LEWIS2)," *Sensors*, vol. 19, no. 7, Art. no. 7, Jan. 2019, doi: 10.3390/s19071549.

[21] D. Maharjan, E. Wyckoff, M. Agüero, S. Martinez, L. Zhou, and F. Moreu, "Monitoring induced floor vibrations: dance performance and bridge engineering," in *Sensors and Smart Structures Technologies for Civil, Mechanical, and Aerospace Systems 2019*, Mar. 2019, vol. 10970, pp. 419–426. doi: 10.1117/12.2515368.

[22] "Adafruit MMA8451 Accelerometer Breakout," *Adafruit Learning System*. https://learn.adafruit.com/adafruit-mma8451-accelerometer-breakout/overview (accessed Oct. 27, 2022).

[23] "Arduino Reference - Arduino Reference." https://www.arduino.cc/reference/en/ (accessed Oct. 27, 2022).

[24] A. Industries, "Adafruit LSM6DSOX 6 DoF Accelerometer and Gyroscope." https://www.adafruit.com/product/4438 (accessed Oct. 28, 2022).

[25] S. F. Barrett, *Arduino Microcontroller Processing for Everyone!* Cham: Springer International Publishing, 2013. doi: 10.1007/978-3-031-79864-1.

[26] T. Dinh, T. Nguyen, H.-P. Phan, V. Dau, D. Dao, and N.-T. Nguyen, "Physical Sensors: Thermal Sensors," in *Encyclopedia of Sensors and Biosensors (First Edition)*, R. Narayan, Ed. Oxford: Elsevier, 2023, pp. 20–33. doi: 10.1016/B978-0-12-822548-6.00052-2.